\documentclass[twocolumn,showpacs,preprintnumbers,superscriptaddress,prb]{revtex4}

\usepackage{graphicx,bm,times,url}
\usepackage{amsmath}
\usepackage{amsthm}
\usepackage{amsfonts}
\usepackage{mathrsfs}
\usepackage{graphics}
\usepackage{float}
\usepackage{datetime}

\graphicspath{{./fits/}{./params_vs_T/}}

\newcommand{\dd}{\mbox{${\rm d}$} {}}
\newcommand{\DD}{\mbox{${\rm D}$} {}}
\newcommand{\II}{\mbox{${\rm I}$} {}}
\newcommand{\EE}[3]{\mbox{$E\left(#1,#2;#3\right)$}}

\newcommand{\LL}{\mbox{${\mathscr L}$} {}}
\newcommand{\Fig}[1]{Fig.~\ref{#1}}

\newcommand{\Tab}[1]{Table~\ref{#1}}

\newcommand{\K}{\,{\rm K}}
\newcommand{\s}{\,{\rm s}}

\begin{document}
\preprint{NORDITA 2012-42}

\title{Excess Wings in Broadband Dielectric Spectroscopy}

\author{Simon Candelaresi}
\affiliation{NORDITA, Royal Institute of Technology and Stockholm University,
Roslagstullsbacken 23, 10691 Stockholm, Sweden}
\affiliation{Department of Astronomy,
Stockholm University, 10691 Stockholm, Sweden}

\author{Rudolf Hilfer}
\affiliation{Insitut f\"ur Computerphysik, Universit\"at Stuttgart,
70569 Stuttgart, Germany}
\affiliation{Institut f\"ur Physik, Universit\"at Mainz, 55099 Mainz, Germany}

\begin{abstract}
Analysis of excess wings in broadband 
dielectric spectroscopy data of glass forming materials 
is found to provide evidence for anomalous time evolutions
and fractional semigroups.
Solutions of fractional
evolution equations in frequency space are used to 
fit dielectric spectroscopy data of glass
forming materials with a range
between 4 and 10 decades in frequency.
We show that with only three parameters
(two relaxation times plus one exponent) 
excellent fits can be obtained
for 5-methyl-2-hexanol and for 
methyl-m-toluate over up to 7 decades.
The traditional Havriliak-Negami fit with three parameters
(two exponents and one relaxation time) fits only
4-5 decades.
Using a second exponent, as in Havriliak-Negami fits,
the $\alpha$-peak and the excess wing
can be modeled perfectly with our theory 
for up to 10 decades
for all materials at all temperatures considered here.
Traditionally this can only be accomplished by combining
two Havriliak-Negami functions with 6 parameters.
The temperature dependent relaxation times are fitted with the
Vogel-Tammann-Fulcher relation which provides the corresponding
Vogel-Fulcher temperatures.
The relaxation times turn out to obey almost perfectly the
Vogel-Tammann-Fulcher law.
Finally we report new and computable expressions of time dependent
relaxation functions corresponding to the frequency dependent 
dielectric susceptibilities.

\end{abstract}

\maketitle

\section{Introduction}
Many physical properties of glass forming liquids (e.g.\ their viscosity)
vary dramatically (often over 15 or more decades) within a
narrow temperature interval \cite{LSBL000}.
This phenomenon is the
glass transition.
The change of physical properties during the
glass transition has not yet been fully understood and 
remains a 
subject of intense investigations
\cite{Lunkenheimer96, Corezzi2, Feldman2, Corezzi, Kalinovskaya, LoidlPC,
LoidlGlycerol, Feldman100, Feldman50, RoskildeNature2008, RoskildeChemPhys2009}.

In this work we study the glass transition by 
observing the dielectric susceptibility.
The dielectric susceptibility quantifies
the response of permanent and
induced dipoles to an applied frequency dependent electric field.
The dielectric loss (resp.\ imaginary part of the
complex dielectric susceptibility) typically shows 
a temperature dependent maximum, the
$\alpha$-peak, at low frequencies.
It is followed at higher frequencies by a so called excess wing
\cite{KS003}.
This excess wing has not yet been understood nor has it been described by
any model with less then 4 fit parameters \cite{KS003}.
Existing theories, such as the mode coupling theory
(see \cite{mode_coupling_gotze98, mode_coupling_Reichman05} and
references therein),
do not allow to fit the excess wing.
Traditional phenomenological
fits of the excess wing employ a superposition of  
two Havriliak-Negami functions \cite{HavriliakNegami66,Corezzi,KS003},
and they need 7 fit parameters to fit a range of 10 decades.

The aim of this work is to 
provide fit functions for excess wings with
only three (four) parameters (one (two) exponent(s) and two relaxation times)
obtained from the
previously
introduced method of fractional time evolution
\cite{hil02c, HilferFracRel2003}, and to apply them to
experimental data exhibiting a clear excess wing on a frequency
range as broad as possible.
Our fitting functions need only 3 (model A) or 4 (model B) parameters,
which is a significant improvement compared to
6 parameters for the superposition
of the Havriliak-Negami and Cole-Cole expression
presently used.
We study the glass forming materials 5-methyl-2-hexanol \cite{Kalinovskaya},
glycerol \cite{LoidlGlycerol} and methyl-m-toluate
\cite{RoskildeChemPhys2009}.

\section{Classical relaxation models}
The Debye relaxation model describes the electric relaxation of
dipoles after switching an applied electric field \cite{fro49}.
The normalized relaxation function $f(t)$,
which corresponds to the polarization,
obeys the Debye law
\begin{eqnarray}\label{eq: Debye}
\left(\tau\frac{\dd}{\dd t}+1\right) f(t) & = & 0,
\end{eqnarray}
with the relaxation time $\tau$ and initial condition $f(0)=1$.
The response function, i.e.\ the dynamical dielectric susceptibility, $\chi$
is related to the relaxation function via \cite{hil02c}
\begin{eqnarray}\label{eq: response relaxation}
\chi(t) = - \frac{\dd}{\dd t} f(t).
\end{eqnarray}
In the following discussions we focus on Laplace transformed
quantities.
We use the Laplace transformation of $f(t)$
\begin{eqnarray}
\LL\{f(t)\}(u) = \int_{0}^{\infty} e^{-ut} f(t) \ \dd t,
\end{eqnarray}
where $u=i\nu$ and $\nu$ is the frequency.
Rewriting equation \eqref{eq: response relaxation} in frequency space and
utilizing $f(0) = 1$ leads to
\begin{eqnarray}
\hat{\chi}(u) & = & \LL\{\chi(t)\}(u) \nonumber\\
 & = & 1 - u\LL\{f(t)\}(u) \label{eq: chi L_f} \nonumber\\
 & = & \frac{1}{1+u\tau},
\end{eqnarray}
the well known Debye susceptibility.

In experiments one measures not the normalized quantity $\hat{\chi}(u)$, 
but instead
\begin{eqnarray}
\varepsilon(u) = (\varepsilon_{0}-\varepsilon_{\infty})\hat{\chi}(u)
+ \varepsilon_{\infty},
\end{eqnarray}
where $\varepsilon_{0}$ and $\varepsilon_{\infty}$ are the dynamical
susceptibilities at low, respectively high frequencies.

The Debye model is not able to describe the experimental data well,
because experimental relaxation peaks are broader and asymmetric.
For this reason other fitting functions were proposed like the
Cole-Cole \cite{cole-cole42}, Cole-Davidson \cite{cole-davidson51}
and Havriliak-Negami \cite{HavriliakNegami66} expressions,
whose normalized forms have typically 2 or 3 parameters 
(see table \ref{tab: fit functions}).
They were introduced purely phenomenologically to fit the data.
This can be considered as a drawback.
These functions with three parameters
are able to fit the data over a range of at most $5$
decades (Havriliak-Negami).
Several copies are commonly superposed to fit a broader range,
e.g.\ Havriliak-Negami plus Cole-Cole, which would result in 6
fit parameters.

\begin{table}[b!]
\caption{
List of traditional fit functions for dielectric spectroscopy
data of glass forming materials.
}
\vspace{12pt}\centerline{\begin{tabular}{lcc}
\hline \hline
 & $\hat{\chi}(u)$ & number of parameters \\
\hline
Cole-Cole & $ 1/(1+(u\tau)^{\alpha})$ & 2 \\
Cole-Davidson & $1/(1+u\tau)^{\alpha}$ & 2 \\
Havriliak-Negami & $1/(1+(u\tau)^{\alpha})^{\gamma}$ & 3 \\
\hline \hline
\end{tabular}}
\label{tab: fit functions}
\end{table}

\section{Fractional relaxation models}
In this work we use a generalized form of the Debye relaxation model
in equation \eqref{eq: Debye}.
It is based on the theory of fractional time evolutions for macroscopic
states of many body systems first proposed in equation (5.5) in \cite{hil93e}
and subsequently elaborated in 
\cite{hil95c,hil95e,hil95f,hil98e,hil00a,hil02a,hil02c,hil03d,hil03e,hil07e,hil11b}.
As discussed in \cite{hil02a,hil02c}
composite fractional time evolutions 
are expected near the glass transition.
Such time evolutions give rise to generalized
Debye laws of the form of model A:
\begin{equation}
\left( \tau_{1}\DD + \tau_{2}^{\alpha}\DD^{\alpha} + 1 \right)
f(t) = 0 \label{eq: fdeq model A} 
\end{equation}
or model B:
\begin{equation}
\left( \tau_{1}\DD + \tau_{1}^{\alpha_{1}}\DD^{\alpha_{1}} +
\tau_{2}^{\alpha_{2}}\DD^{\alpha_{2}} + 1 \right) f(t) = 0,
\label{eq: fdeq model B} 
\end{equation}
where the parameters obey $0 < \alpha, \alpha_{1}, \alpha_{2} < 1$, 
$\alpha_{1} > \alpha_{2}$
and the relaxation times $\tau_1,\tau_2>0$ are positive.
Here the symbols
$\tau_{1}\DD + \tau_{2}^{\alpha}\DD^{\alpha}$,
respectively
$\tau_{1}\DD + \tau_{1}^{\alpha_{1}}\DD^{\alpha_{1}} +
\tau_{2}^{\alpha_{2}}\DD^{\alpha_{2}}$
are the infinitesimal generators 
of composite fractional semigroups with
$\DD^{\alpha}$
being a generalized fractional Riemann-Liouville
derivative of order $\alpha$ and almost any type
\cite{hil00a,hil09d}.
If $\DD^{\nu}$ represents a classical fractional
Riemann-Liouville derivative of order $\nu$
then its definition reads
(with $\nu \in \mathbb{R}^{+}$)
\begin{eqnarray}
\DD^{\nu} f(t)  & = &
\DD^{\lceil \nu \rceil} \II^{\mu} f(t) \\
 & = & \frac{1}{\Gamma(\mu)} \DD^{\lceil \nu \rceil}
\int\limits_{0}^{t}(t-\xi)^{\mu-1} f(\xi) \ \dd \xi, \\
& & \mu + \nu = \lceil \nu \rceil, \quad t > 0, \nonumber
\end{eqnarray}
where $\lceil \nu \rceil$ is the smallest integer greater or equal
$\nu$, $\Gamma$ the gamma function and
$\DD^{\lceil \nu \rceil} = \dd^{\lceil \nu \rceil}/\dd t^{\lceil \nu \rceil}$.

The Laplace transform of the fractional Riemann-Liouville derivative is
\cite{MillerRoss}
\begin{eqnarray}\label{eq: Laplace fractional}
\LL\{\DD^{\nu} f(t)\}(u) & = & u^{\nu}\LL\{f(t)\}(u) \nonumber \\
 & & - \sum_{k=1}^{\lceil \nu \rceil} u^{k-1} \left. \DD^{\nu-k}
 f(t) \right|_{t=0}.
\end{eqnarray}

With these definitions the Laplace transformation of equations
\eqref{eq: fdeq model A} and \eqref{eq: fdeq model B}
gives with relation \eqref{eq: response relaxation} 
the normalized dielectric susceptibilities of model A
\begin{equation}
\hat{\chi}_{\rm A}(u) =
\frac{\displaystyle{1+\tau_{2}^{\alpha}u^{\alpha}}}
{\displaystyle{\tau_{1}u+\tau_{2}^{\alpha}u^{\alpha}+1}} \label{eq: model A}
\end{equation}
and model B
\begin{equation}
\hat{\chi}_{\rm B}(u) =
\frac{\displaystyle{1+\tau_{1}^{\alpha_{1}}u^{\alpha_{1}} +
\tau_{2}^{\alpha_{2}}u^{\alpha_{2}}}}
{\displaystyle{\tau_{1}u+\tau_{1}^{\alpha_{1}}u^{\alpha_{1}} +
\tau_{2}^{\alpha_{2}}u^{\alpha_{2}}+1}} \label{eq: model B}.
\end{equation}
These results apply also for other types of generalized
Riemann-Liouville fractional derivatives 
introduced in \cite{hil02a,hil02c}.

The functions from equations \eqref{eq: model A} and \eqref{eq: model B}
are used to fit the
dielectric spectroscopy data of 5-methyl-2-hexanol,
glycerol and methyl-m-toluate.
Real and imaginary part are fitted simultaneously with the parameters
$\alpha$, $\alpha_{1}$, $\alpha_{2}$, $\tau_{1}$ and $\tau_{2}$.

Additionally we fit the temperature dependent relaxation times
$\tau_{1}$ and $\tau_{2}$ with the
Vogel-Tammann-Fulcher function
\begin{eqnarray} \label{eq: vtf}
\tau = \tau_{0} \exp\left(\frac{DT_{\rm VF}}{T - T_{\rm VF}}\right),
\end{eqnarray}
where $T$ is the absolute temperature, $\tau_{0}$ a material
parameter, $D$ the fragility and $T_{\rm VF}$ the Vogel-Fulcher temperature.
The fit parameters are $\tau_{0}$, $D$ and $T_{\rm VF}$.

\section{Results}

\begin{figure}[t!]
\begin{center}
\includegraphics[height=\columnwidth]{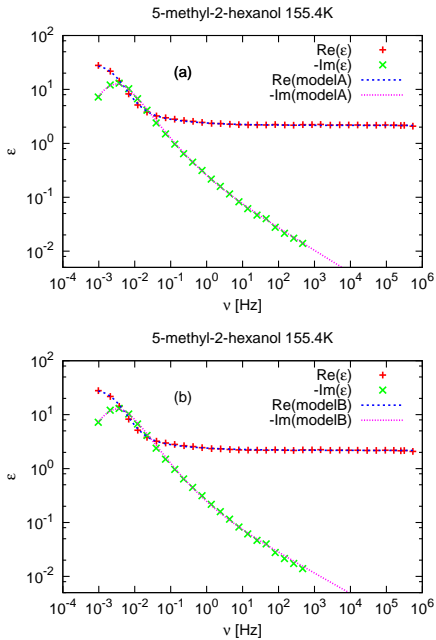}
\end{center}
\caption{
(Color online)
Simultaneous fits of real and imaginary part with model A (upper figure)
and model B (lower figure) for 5-methyl-2-hexanol at $155.4\K$.
Both models show an excellent fitting capability.
The data are from \cite{Kalinovskaya}.
}
\label{fig: fit alcohol carbonate}
\end{figure}

\begin{figure}[t!]
\begin{center}
\includegraphics[height=\columnwidth]{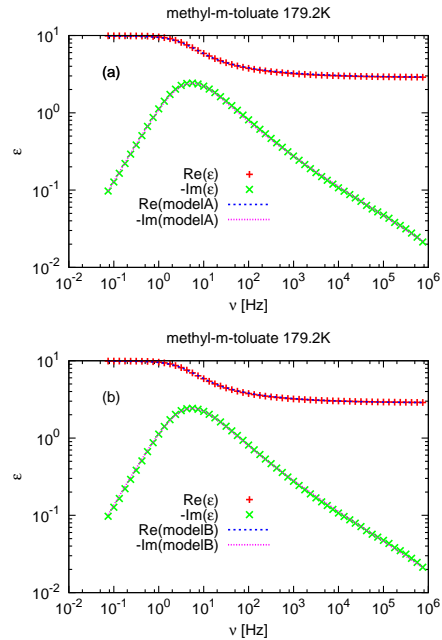}
\end{center}
\caption{
(Color online)
Simultaneous fits of real and imaginary part with model A (upper figure) and
model B (lower figure) for methyl-m-toluate at $179.2\K$.
Model A can fit the data over the whole spectral range of 7 decades, which
is more then with Havriliak-Negami which uses the same number
of fit parameters.
With this data we obtain the broadest fit with model A.
The data are from \cite{RoskildeChemPhys2009}.
}
\label{fig: fit methyl-m-toluate}
\end{figure}

\begin{figure}[t!]
\begin{center}
\includegraphics[height=\columnwidth]{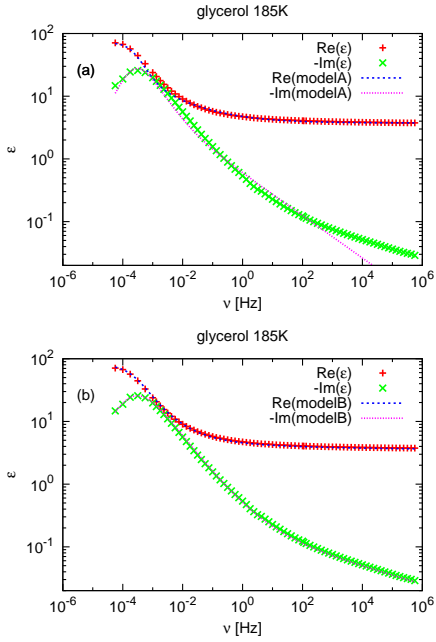}
\end{center}
\caption{
(Color online)
Simultaneous fits of real and imaginary part with model A (upper figure) and
model B (lower figure) for glycerol at $185\K$.
While model A is not able to fit the data well, model B
still gives an excellent fit over 10 decades.
The data are from \cite{LoidlGlycerol}.
}
\label{fig: fit glycerol}
\end{figure}

Model A fits the 5-methyl-2-hexanol
(\Fig{fig: fit alcohol carbonate}) and
methyl-m-toluate (\Fig{fig: fit methyl-m-toluate}) data remarkably well
for 5, respectively 7 orders of magnitude, where
both the $\alpha$-peak and the excess wing can be fitted
simultaneously with only three parameters.
Model A is better suited then the Havriliak-Negami model which
only fits reasonably well for up to four orders of magnitudes
for these materials.
This improvement is due to the positive curvature 
of the function in \eqref{eq: model A} at frequencies above 
the $\alpha$-peak.
Sometimes this curvature poses also the main difficulty 
when fitting with model A.
An example is glycerol as seen in the upper part of \Fig{fig: fit glycerol}.
While it is easy to fit closely the the $\alpha$-peak
it is more difficult to simultaneously fit the excess wing.

Model B can fit the data much better then model A, which is not a surprise
since it comes with one more parameter.
Nevertheless, it is remarkable that it can fit a range of up to 10 orders of
magnitude with little deviation from the data points.
We believe that this model can be used to fit over some more orders
of magnitude, but at this time there is no experimental data available
which covers a broader range.

\section{Temperature dependence of the parameters}
\label{sec: Temperature dependence of the parameters}
Because the data have been fitted at different temperatures we are able
to observe the temperature dependence of the fitting parameters.
For $\tau_{1}$ and $\tau_{2}$ we perform Vogel-Tammann-Fulcher fits
provided by equation \eqref{eq: vtf}.
Note that in our notation $\varepsilon = \varepsilon' - i\varepsilon''$.
From the fits we obtain the Vogel-Fulcher temperatures $T_{\rm VF 1}$ and
$T_{\rm VF 2}$ as well as the fragility parameters $D_{1}$ and
$D_{2}$ for the relaxation times $\tau_{1}$ and $\tau_{2}$
for model A and model B (see \Tab{tab: fitting parameters}).
    
\begin{table*}[b!] 
\caption{
List of the fit parameters $T_{\rm VF}$, $D$ and $\tau_{0}$
for various materials.
}
\vspace{12pt}\centerline{\begin{tabular}{lccccccc}
\hline \hline
material & model & $T_{\rm VF 1}$ & $T_{\rm VF 2}$ & $D_{1}$ & $D_{2}$ &
$\tau_{01}$ & $\tau_{02}$ \\
\hline
5-methyl-2-hexanol & A & $89.2\K$ & $92.1\K$ & $25.5$ & $22.1$
 & $3.53\times10^{-13}\s$ & $7\times10^{-14}\s$ \\
5-methyl-2-hexanol & B & $88.3\K$ & $101.6\K$ & $26.3$ & $15.3$
 & $5.21\times10^{-13}\s$ & $5.77\times10^{-12}\s$ \\
glycerol & A & $127.8\K$ & $131.5\K$ & $17.1$ & $14.9$
 & $3.7\times10^{-14}\s$ & $3.23\times10^{-14}\s$ \\
glycerol & B & $152.7\K$ & $137.8\K$ & $6.68$ & $11.9$
 & $2.9\times10^{-10}\s$ & $2.15\times10^{-13}\s$ \\
methyl-m-toluate & A & $71.2\K$ & $71.2\K$ & $93.2$ & $93.2$
 & $4.35\times10^{-28}\s$ & $1.54\times10^{-28}\s$ \\
methyl-m-toluate & B & $67.2\K$ & $85.6\K$ & $102.2$ & $53.6$
 & $9.0\times10^{-28}\s$ & $1.13\times10^{-22}\s$ \\
\hline \hline
\end{tabular}}
\label{tab: fitting parameters}
\end{table*}

\begin{figure}[t!]
\begin{center}
\includegraphics[width=\columnwidth]
{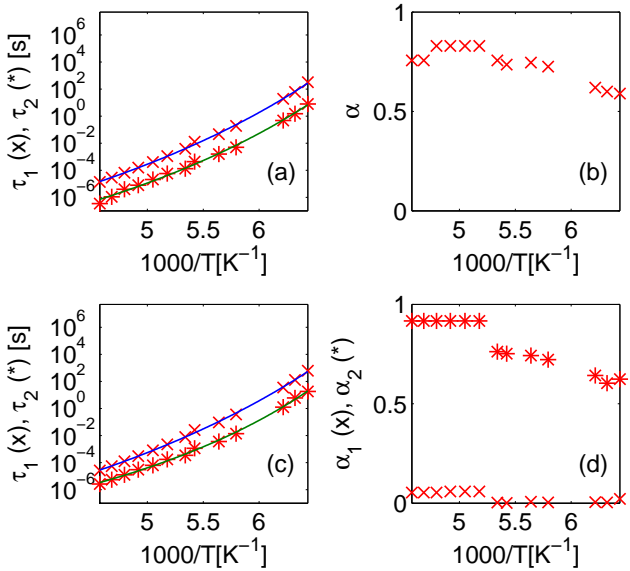}
\end{center}
\caption{
(Color online)
Temperature dependence of the fitting parameters for
5-methyl-2-hexanol for model A
(upper panels) and model B (bottom panels) together with
Vogel-Tammann-Fulcher fits.
}
\label{fig: Kalinkovskaya}
\end{figure}

\begin{figure}[t!]
\begin{center}
\includegraphics[width=\columnwidth]
{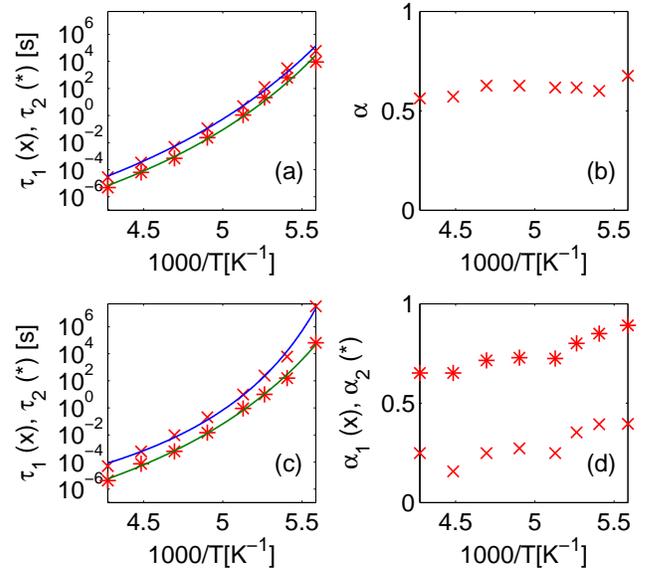}
\end{center}
\caption{
(Color online)
Temperature dependence of the fitting parameters for
glycerol for model A
(upper panels) and model B (bottom panels) together with
Vogel-Tammann-Fulcher fits.
}
\label{fig: Loidl glycerol}
\end{figure}

\begin{figure}[t!]
\begin{center}
\includegraphics[width=\columnwidth]
{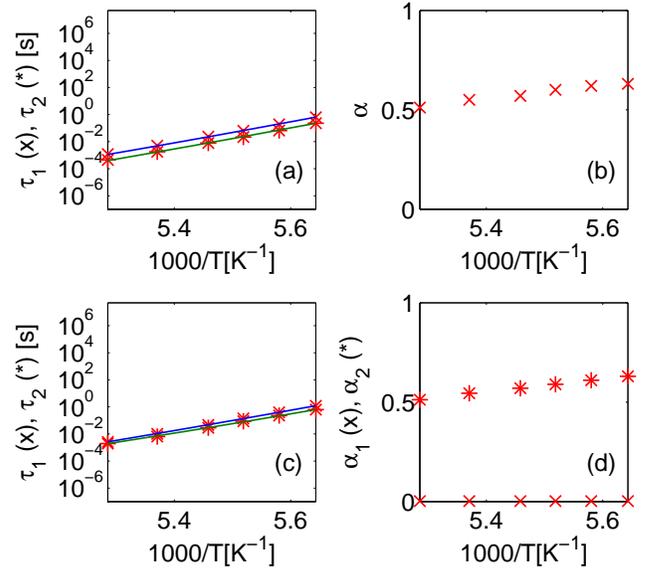}
\end{center}
\caption{
(Color online)
Temperature dependence of the fitting parameters for
methyl-m-toluate for model A
(upper panels) and model B (bottom panels) together with
Vogel-Tammann-Fulcher fits.
}
\label{fig: Roskilde MMT}
\end{figure}

For all fits we see a temperature dependence of the 
relaxation times $\tau_{1}$ and
$\tau_{2}$ (\Fig{fig: Kalinkovskaya} - \Fig{fig: Roskilde MMT})
that follows the Vogel-Tammann-Fulcher fitting function
remarkably well.
The relaxation times also show a clear downward trend as the
temperature increases, which confirms that $\tau$, $\tau_{1}$ and
$\tau_{2}$ are physically
meaningful and can be interpreted as relaxation times
even tough they appear with a non-integer power in equations
\eqref{eq: model A} and \eqref{eq: model B}.

The parameters $\alpha$, $\alpha_{1}$ and $\alpha_{2}$ also
show a temperature dependence.
In the case of 5-methyl-2-hexanol (\Fig{fig: Kalinkovskaya})
there is an increase of $\alpha$ with temperature until a plateau
near $\alpha = 1$ is reached.
This effect comes from the decreasing slope of the excess wing with
increasing temperature.
In the fitting function of model A this behavior can be achieved by
increasing $\alpha$.
For the same material there is an apparent increase of $\alpha_{2}$ between
$154\K$ ($6.49\K^{-1}$) and $287\K$ ($3.48\K^{-1}$)
which has the same origin as the increase in $\alpha$ in
model A.
By increasing $\alpha_{2}$ the excess wing becomes less steep.
The plateau at $190\K$ ($5.26\K^{-1}$) and above comes from the fact that the fits at those
temperatures are done mainly for the $\alpha$-peak since the excess wing
is not visible.

For glycerol and methyl-m-toluate there is also a clear temperature dependence
of $\alpha$, $\alpha_{1}$ and $\alpha_{2}$
(\Fig{fig: Loidl glycerol} and \Fig{fig: Roskilde MMT}).
The trend is however reversed in comparison to 5-methyl-2-hexanol.
This comes from the increasing slope of the excess wing with
increasing temperature.
This behavior can be achieved in the fit functions by decreasing $\alpha$,
respectively $\alpha_{2}$.

\section{Representation of the solutions as functions of time}

In \cite{CandelaresiMaster08}
we obtained the analytical solution of a fractional differential equation
of rational order, which we use to analyze our fitting results
for model A and model B.
For a general solution of equations \eqref{eq: model A} and \eqref{eq: model B}
with arbitrary real $\alpha_i$ see \cite{hil09d}.
The restriction to rational $\alpha_i$ is not a drawback, since we can
approximate $\alpha_{1}$ and $\alpha_{2}$ by a rational value on a grid between 0 and 1.
This number of grid points is chosen to be 20, which keeps computation
times reasonably limited as the computing time increases quadratically
with the lowest common denominator of $\alpha$ with 1.

The solution for $f(t)$ for model B is a sum
of Mittag-Leffler type functions:
\begin{eqnarray}\label{eq: solution fdeq}
f(t) = \sum_{j=1}^{N}B_{j} \sum_{k=0}^{N-1} c_{j}^{N-k-1} \EE{-k/N}{c_{j}^{N}}{t},
\end{eqnarray}
where $N$ is the smallest number for which both $\alpha_{1}N$ and
$\alpha_{2}N$ are integers.
The coefficients $c_{j}$ are the zeros of the characteristic polynomial
\begin{eqnarray} \label{eq: indicial polynomial}
c^{N} + \tau_{1}^{\alpha_{1}}c^{\alpha_{1}N}
+ \tau_{2}^{\alpha_{2}}c^{\alpha_{2}N} + 1 = 0,
\end{eqnarray}
the function $\EE{\nu}{a}{t}$ is defined as \cite{MillerRoss}
\begin{eqnarray}
\EE{\nu}{a}{t} = t^{\nu} \sum_{k=0}^{\infty} \frac{(at)^{k}}{\Gamma(\nu+k+1)}.
\end{eqnarray}
The coefficients $B_{j}$ are the solutions of 
the linear system of equations
\begin{eqnarray}
 &\sum\limits_{k=1}^{N} c_{k}^{i} B_{k} = 0, \label{eq: lse1} \\
 &0 \le i \le N-\alpha_{1} N - 1 \nonumber \\
 &\sum\limits_{k=1}^{N} (c_{k}^{i} + 
 \tau_{1}^{\alpha_{1}} c_{k}^{i-N+\alpha_{1} N}) B_{k} = 0, \label{eq: lse2} \\
 &N-\alpha_{1} N \le i \le N-\alpha_{2} N -1 \nonumber \\
 &\sum\limits_{k=1}^{N} (c_{k}^{i} + \tau_{1}^{\alpha_{1}} c_{k}^{i-N+\alpha_{1} N}
+ \tau_{2}^{\alpha_{2}} c_{k}^{i-N+\alpha_{2} N}) B_{k} = 0, \label{eq: lse3} \\
 &N-\alpha_{2} N \le i \le N-2. \nonumber
\end{eqnarray}
This solution is only valid if all the roots $c_{j}$
of the characteristic polynomial in \eqref{eq: indicial polynomial}
are distinct, which is checked in the computations.
Since the linear system of equations \eqref{eq: lse1}-\eqref{eq: lse3} is
underdetermined we choose one fundamental solution for $\{B_{j}\}$ and a
multiplication factor for $f(t)$ such that $f(0) = 1$.

\begin{figure}[t!]
\begin{center}
\includegraphics[width=\columnwidth]
{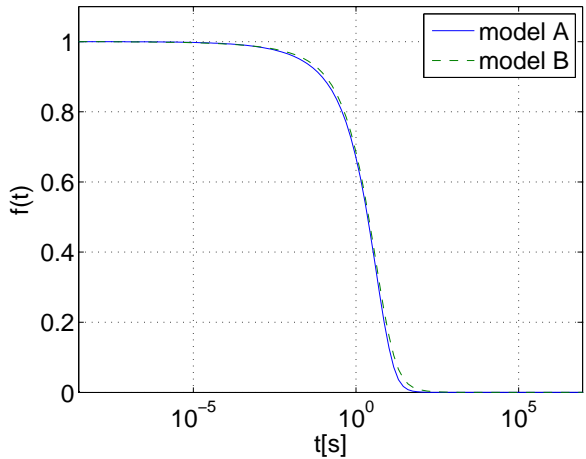}
\caption{
(Color online)
The solutions of the fractional initial value problems
\eqref{eq: fdeq model A} and \eqref{eq: fdeq model B} with
$f(0) = 1$ using the fit parameters for model A and model B for
glycerol at $195\K$.
}
\label{fig: time dependence}
\end{center}
\end{figure}

The analytical solutions are plotted
for glycerol at $195\K$ (\Fig{fig: time dependence}).
The fitting values for $\tau$ for model A are $\tau_{1} = 4.991\s$
and $\tau_{2} = 1.089\s$.
Both values lie in the time interval where the relaxation occurs,
which confirms the interpretation of these fitting parameters as
relaxation times.
For model B the fitted times are $\tau_{1} = 9.729\s$ and $\tau_{2} = 0.92\s$.
So $\tau_{2}$ marks the onset of the relaxation and $\tau_{1}$ the
end.

We note that the fractional derivatives appearing in
the initial value problem \eqref{eq: fdeq model B} 
can be generalized to fractional derivatives
of arbitrary type $\beta$ introduced in 
\cite{hil00a}
and defined as
\begin{eqnarray}
\label{eq:GRLFD}
 & \DD^{\nu,\beta} f(t) =
\II^{(1-\beta)(\lceil\nu\rceil - \nu)}
\DD^{\lceil\nu\rceil}
\II^{\beta(\lceil\nu\rceil - \nu)} f(t), \\
 & 0 \le \beta \le 1. \nonumber
\end{eqnarray}
For the case $\beta = 1$ it reduces to the Riemann-Liouville
fractional derivative, while for $\beta = 0$ to the Caputo-type
derivative \cite{Caputo67}.
Because
\begin{eqnarray}
 & \DD^{\nu,\beta} \EE{\mu}{a}{t} = \DD^{\nu,\gamma} \EE{\mu}{a}{t}, \\
 & 0 \le \beta, \gamma \le 1, \quad \mu > -1, \quad \nu \ge 0, \nonumber
\end{eqnarray}
the solution of our initial value problem does not change by replacing
the Riemann-Liouville fractional derivatives with these generalized
Riemann-Liouville fractional derivatives of type $\beta$.

\section{Conclusions}
The two fractional relaxation models (model A and model B) are shown
to fit well dielectric spectroscopy data for various glass forming 
materials over
a range of at least 5 orders of magnitude for model A and 10 orders
of magnitude for model B.
For 5-methyl-2-hexanol and methyl-m-toluate model A can fit data
over a range of 6 and 7 orders of magnitude.
This is a significant improvement over conventional fitting formulae like
Havriliak-Negami and Cole-Cole which need to be superimposed in order
to fit data at the same range.
Conventional fitting formulae 
require 6 parameters in contrast to 3 for our model A
and 4 for our model B.
Another advantage of the fractional relaxation models is that
the fitting formulas follow from an underlying
general theory based on composition of fractional semigroups
and their infinitesimal generators,
which is not the case for Havriliak-Negami or Cole-Cole
functions.

\section{Acknowledgements}
We thank Sebastian Schmiescheck for converting data into a workable format.

\bibliographystyle{ieeetr}
\bibliography{references}

\begin{thebibliography}{10}

\bibitem{LSBL000}
P.~{Lunkenheimer}, U.~{Schneider}, R.~{Brand}, and A.~{Loid}, ``{Glassy
  dynamics},'' {\em Contemporary Physics}, vol.~41, pp.~15--36, 2000.

\bibitem{Lunkenheimer96}
P.~Lunkenheimer, A.~Pimenov, B.~Schiener, R.~B{\"o}hmer, and A.~Loidl,
  ``High-frequency dielectric spectroscopy on glycerol,'' {\em EPL (Europhysics
  Letters)}, vol.~33, no.~8, p.~611, 1996.

\bibitem{Corezzi2}
S.~Corezzi, S.~Capaccioli, G.~Gallone, M.~Lucchesi, and P.~A. Rolla, ``Dynamics
  of a glass-forming triepoxide studied by dielectric spectroscopy,'' {\em
  Journal of Physics: Condensed Matter}, vol.~11, no.~50, pp.~10297--10314,
  1999.

\bibitem{Feldman2}
R.~Behrends, K.~Fuchs, U.~Kaatze, Y.~Hayashi, and Y.~Feldman, ``Dielectric
  properties of glycerol/water mixtures at temperatures between 10 and 50
  [degree]c,'' {\em The Journal of Chemical Physics}, vol.~124, no.~14,
  p.~144512, 2006.

\bibitem{Corezzi}
S.~Corezzi, M.~Beiner, H.~Huth, K.~Schr\"{o}ter, S.~Capaccioli, R.~Casalini,
  D.~Fioretto, and E.~Donth, ``Two crossover regions in the dynamics of glass
  forming epoxy resins,'' {\em The Journal of Chemical Physics}, vol.~117,
  no.~5, pp.~2435--2448, 2002.

\bibitem{Kalinovskaya}
O.~E. Kalinovskaya and J.~K. Vij, ``The exponential dielectric relaxation
  dynamics in a secondary alcohol's supercooled liquid and glassy states,''
  {\em The Journal of Chemical Physics}, vol.~112, no.~7, pp.~3262--3266, 2000.

\bibitem{LoidlPC}
U.~Schneider, P.~Lunkenheimer, R.~Brand, and A.~Loidl, ``Broadband dielectric
  spectroscopy on glass-forming propylene carbonate,'' {\em Phys. Rev. E},
  vol.~59, no.~6, pp.~6924--6936, 1999.

\bibitem{LoidlGlycerol}
P.~Lunkenheimer, A.~Pimenov, M.~Dressel, Y.~G. Goncharov, R.~B{\"o}hmer, and
  A.~Loidl, ``Fast dynamics of glass-forming glycerol studied by dielectric
  spectroscopy,'' {\em Phys. Rev. Lett.}, vol.~77, no.~2, pp.~318--321, 1996.

\bibitem{Feldman100}
A.~Puzenko, Y.~Hayashi, Y.~Ryabov, I.~Balin, Y.~Feldman, U.~Kaatze, and
  R.~Behrends, ``Relaxation dynamics in glycerol-water mixtures: I.
  glycerol-rich mixtures,'' {\em Journal of Physical Chemistry B}, vol.~109,
  no.~12, pp.~6031--6035, 2005.

\bibitem{Feldman50}
Y.~Hayashi, A.~Puzenko, I.~Balin, Y.~Ryabov, and Y.~Feldman, ``Relaxation
  dynamics in glycerol-water mixtures. 2. mesoscopic feature in water rich
  mixtures,'' {\em Journal of Physical Chemistry B}, vol.~109, no.~18,
  pp.~9174--9177, 2005.

\bibitem{RoskildeNature2008}
T.~{Hecksher}, A.~I. {Nielsen}, N.~B. {Olsen}, and J.~C. {Dyre}, ``Little
  evidence for dynamic divergences in ultraviscous molecular liquids,'' {\em
  Nat. Phys.}, vol.~4, pp.~737 -- 741, 2008.

\bibitem{RoskildeChemPhys2009}
A.~I. Nielsen, T.~Christensen, B.~Jakobsen, K.~Niss, N.~B. Olsen, R.~Richert,
  and J.~C. Dyre, ``Prevalence of approximate sqrt(t) relaxation for the
  dielectric alpha process in viscous organic liquids,'' {\em The Journal of
  Chemical Physics}, vol.~130, no.~15, p.~154508, 2009.

\bibitem{KS003}
F.~Kremer and A.~Sch{\"o}nhals, {\em Broadband Dielectric Spectroscopy}.
\newblock Berlin: Springer Verlag, 2003.

\bibitem{mode_coupling_gotze98}
W.~G{\"o}tze, ``The essentials of the mode-coupling theory for glassy
  dynamics,'' {\em Condensed Matter Physics}, vol.~1, no.~4, p.~873, 1998.

\bibitem{mode_coupling_Reichman05}
D.~R. Reichman and P.~Charbonneau, ``Mode-coupling theory,'' {\em Journal of
  Statistical Mechanics: Theory and Experiment}, vol.~2005, no.~05, p.~P05013,
  2005.

\bibitem{HavriliakNegami66}
S.~Havriliak and S.~Negami, ``A complex plane analysis of $\alpha$-dispersions
  in some polymer systems,'' {\em Journal of Polymer Science Part C: Polymer
  Symposia}, vol.~14, no.~1, pp.~99--117, 1966.

\bibitem{hil02c}
R.~Hilfer, ``Experimental evidence for fractional time evolution in glass
  forming materials,'' {\em Chemical Physics}, vol.~284, no.~1–2, pp.~399 --
  408, 2002.

\bibitem{HilferFracRel2003}
R.~Hilfer, ``On fractional relaxation,'' {\em Fractals}, vol.~11, pp.~251--257,
  2003.

\bibitem{fro49}
H.~Fr{\"o}hlich, {\em Theory of Dielectrics: Dielectric Constant and Dielectric
  Loss}.
\newblock Oxford University Press, 1949.

\bibitem{cole-cole42}
K.~S. Cole and R.~H. Cole, ``Dispersion and absorption in dielectrics i.
  alternating current characteristics,'' {\em The Journal of Chemical Physics},
  vol.~9, no.~4, pp.~341--351, 1941.

\bibitem{cole-davidson51}
D.~W. Davidson and R.~H. Cole, ``Dielectric relaxation in glycerol, propylene
  glycol, and n-propanol,'' {\em The Journal of Chemical Physics}, vol.~19,
  no.~12, pp.~1484--1490, 1951.

\bibitem{hil93e}
R.~{Hilfer}, ``{Classification theory for anequilibrium phase transitions},''
  {\em Phys. Rev. E}, vol.~48, pp.~2466--2475, 1993.

\bibitem{hil95c}
R.~{Hilfer}, ``{Fractional dynamics, irreversibility and ergodicity
  breaking},'' {\em Chaos Solitons {\&} Fractals}, vol.~5, pp.~1475--1484,
  1995.

\bibitem{hil95e}
R.~{Hilfer}, ``{Foundations of fractional dynamics},'' {\em Fractals}, vol.~3,
  pp.~549--556, 1995.

\bibitem{hil95f}
R.~{Hilfer}, ``{An extension of the dynamical foundation for the statistical
  equilibrium concept},'' {\em Physica A}, vol.~221, pp.~89--96, 1995.

\bibitem{hil98e}
R.~Hilfer, {\em Applications of Fractional Calculus in Physics}.
\newblock Singapore: World Scientific Publ. Co., 2000.

\bibitem{hil00a}
R.~Hilfer, ``Fractional time evolution,'' in {\em Applications of Fractional
  Calculus in Physics} (R.~Hilfer, ed.), p.~87, Singapore: World Scientific
  Publ. Co., 2000.

\bibitem{hil02a}
R.~{Hilfer}, ``{Fitting the excess wing in the dielectric {$\alpha$}-relaxation
  of propylene carbonate},'' {\em Journal of Physics: Condensed Matter},
  vol.~14, pp.~2297--2301, 2002.

\bibitem{hil03d}
R.~Hilfer, ``Remarks on fractional time,'' in {\em Time, Quantum and
  Information} (L.~Castell and O.~Ischebeck, eds.), p.~235, Berlin: Springer
  Verlag, 2003.

\bibitem{hil03e}
R.~{Hilfer}, ``{On fractional diffusion and continuous time random walks},''
  {\em Physica A}, vol.~329, pp.~35--40, 2003.

\bibitem{hil07e}
R.~Hilfer, ``Threefold introduction to fractional derivatives,'' in {\em
  Anomalous Transport: Foundations and Applications} (R.~Klages, G.~Radons, and
  I.~Sokolov, eds.), pp.~17--73, Wiley-VCH, 2008.

\bibitem{hil11b}
R.~Hilfer, ``Foundations of fractional dynamics: A short account,'' in {\em
  Fractional Dynamics: Recent Advances} (J.~Klafter, S.~Lim, and R.~Metzler,
  eds.), p.~207, Singapore: World Scientific Publ. Co., 2001.

\bibitem{hil09d}
R.~{Hilfer}, Y.~{Luchko}, and Z.~{Tomovski}, ``{Operational method for the
  solution of fractional differential equations with generalized
  {R}iemann-{L}iouville fractional derivatives},'' {\em Fractional Calculus and
  Applied Analysis}, vol.~12, p.~299, 2009.

\bibitem{MillerRoss}
K.~S. Miller and B.~Ross, {\em An Introduction to fractional Calculus and
  fractional Differential Equations}.
\newblock Wiley-Interscience, 1993.

\bibitem{CandelaresiMaster08}
S.~Candelaresi, ``Fraktionale ans{\"a}tze in dielektrischer
  breitbandspektroskopie,'' Master's thesis, 2008.

\bibitem{Caputo67}
M.~Caputo, ``Linear models of dissipation whose q is almost frequency
  independent-ii,'' {\em Geophysical Journal of the Royal Astronomical
  Society}, vol.~13, no.~5, pp.~529--539, 1967.

\end{thebibliography}

\end{document}